\documentclass[aps,prb,amsmath,amssymb,superscriptaddress,twocolumn,color,epsfig,graphicx,bm,floatfix]{revtex4-1}

\usepackage{graphicx}
\usepackage[caption=false]{subfig}
\usepackage{epstopdf}
\usepackage{tabularx}
\usepackage[super]{nth}
\usepackage[caption=false]{subfig}
\usepackage{amsmath}    
\usepackage{mathtools}
\usepackage{siunitx}
\usepackage{multirow}
\usepackage{array}
\usepackage[version=4]{mhchem}
\newcolumntype{K}[1]{>{\centering\arraybackslash}p{#1}}

\usepackage{verbatim}   
\usepackage{color}      
\usepackage{hyperref}   
\begin{document}

\title{Dense superconducting phases of copper-bismuth at high pressure} 

\author{Maximilian Amsler}
\affiliation{Department of Materials Science and Engineering, Northwestern University, Evanston, Illinois 60208, USA}
\author{Chris Wolverton}
\email{c-wolverton@northwestern.edu}
\affiliation{Department of Materials Science and Engineering, Northwestern University, Evanston, Illinois 60208, USA}

\date{\today}

\begin{abstract}
Although copper and bismuth do not form any compounds at ambient conditions, two intermetallics, \ce{CuBi} and \ce{Cu11Bi7}, were recently synthesized at high pressures. Here we report on the discovery of additional copper-bismuth phases at elevated pressures with high-densities from \textit{ab initio} calculations. In particular, a \ce{Cu2Bi} compound is found to be thermodynamically stable at pressures above  59~GPa, crystallizing in the cubic Laves structure. In strong contrast to \ce{Cu11Bi7} and CuBi, cubic \ce{Cu2Bi} does not exhibit any voids or channels. Since the bismuth lone pairs in cubic \ce{Cu2Bi} are stereochemically inactive, the constituent elements can be closely packed and a high density of 10.52~g/cm$^{3}$ at 0~GPa is achieved. The moderate electron-phonon coupling of  $\lambda=0.68$ leads to a superconducting temperature of 2~K, which exceeds the values observed both in \ce{Cu11Bi7} and CuBi, as well as in elemental Cu and Bi . 
\end{abstract}

\maketitle


Intermetallic compounds have been the subject of intense research not only as strengthening precipitates in a wide variety of structural alloys~\cite{nie_precipitation_2012,wang_precipitates_2005}, but also for other industrially relevant applications due to their compelling physical properties, ranging from strong permanent magnetism~\cite{fidler_rare-earth_2007} (e.g. \ce{Nd2Fe14B}~\cite{Croat1984,Sagawa1984} and \ce{SmCo5}~\cite{Benz1970}) and superconductivity (e.g. \ce{FeBi2}~\cite{amsler_prediction_2017,walsh_discovery_2016}, Ca$_{11}$Bi$_{10-x}$~\cite{sturza_superconductivity_2014,dong_rich_2015}, \ce{NiBi}~\cite{Haegg1929}, \ce{NiBi3}~\cite{Glagoleva1954,Ruck2006}, and \ce{CoBi3}~\cite{Matthias1961,Schwarz2013,Tence2014}) to their promising conversion efficiencies in thermoelectric materials~\cite{sun_huge_2009} (e.g. \ce{FeAs2}~\cite{Buerger1932},  \ce{FeSb2}~\cite{Hagg1928} and \ce{Bi2Te3}~\cite{Satterthwaite1957}). Recently, binary intermetallics have attracted increasing attention in systems that exhibit severe immiscibility at ambient conditions, but form compounds once exposed to sufficiently high pressures. Although many such high-pressure compounds in immiscible systems had been synthesized in the early 1960s by Matthias~\textit{et al.}~\cite{Matthias1961}, merely a few of them have been so far fully characterized with respect to their crystal structure, composition and properties. Only through the recent advances in experimental high-pressure techniques together with computational methods with high predictive accuracy~\cite{oganov_modern_2010,zhang_materials_2017} has it become possible to explore and study such high pressure phases with increasing detail and at a much larger scale.

One such ambient-immiscible system that has proven to be especially rich in high-pressure compounds is Cu--Bi, where two novel phases were discovered since 2016~\cite{Clarke2016,guo_weak_2017,clarke_creating_2017}. \ce{Cu11Bi7} was synthesized at 6~GPa, and crystallizes in a new hexagonal structure related to the NiAs  type~\cite{Clarke2016}. It can be recovered to ambient pressure and is a superconductor with a transition temperature of $T_c=1.36$~K. On the other hand, CuBi was synthesized at 5~GPa and 720~$^\circ$C~\cite{guo_weak_2017,clarke_creating_2017} in an orthorhombic structure, and has a slightly lower value of $T_c= 1.3$~K. A peculiar structural feature that both phases have in common is the formation of porous voids in their crystal structure. \ce{Cu11Bi7} exhibits empty channels running along the c-axis of the hexagonal cell, while CuBi contains 2-dimensional empty layers. In fact, we recently demonstrated that CuBi is composed of superconducting 2D sheets, so-called cubine, that are held together through weak van der Waals forces~\cite{amsler_cubine_2017}.

\begin{figure}[b]
\includegraphics[width=0.95\columnwidth]{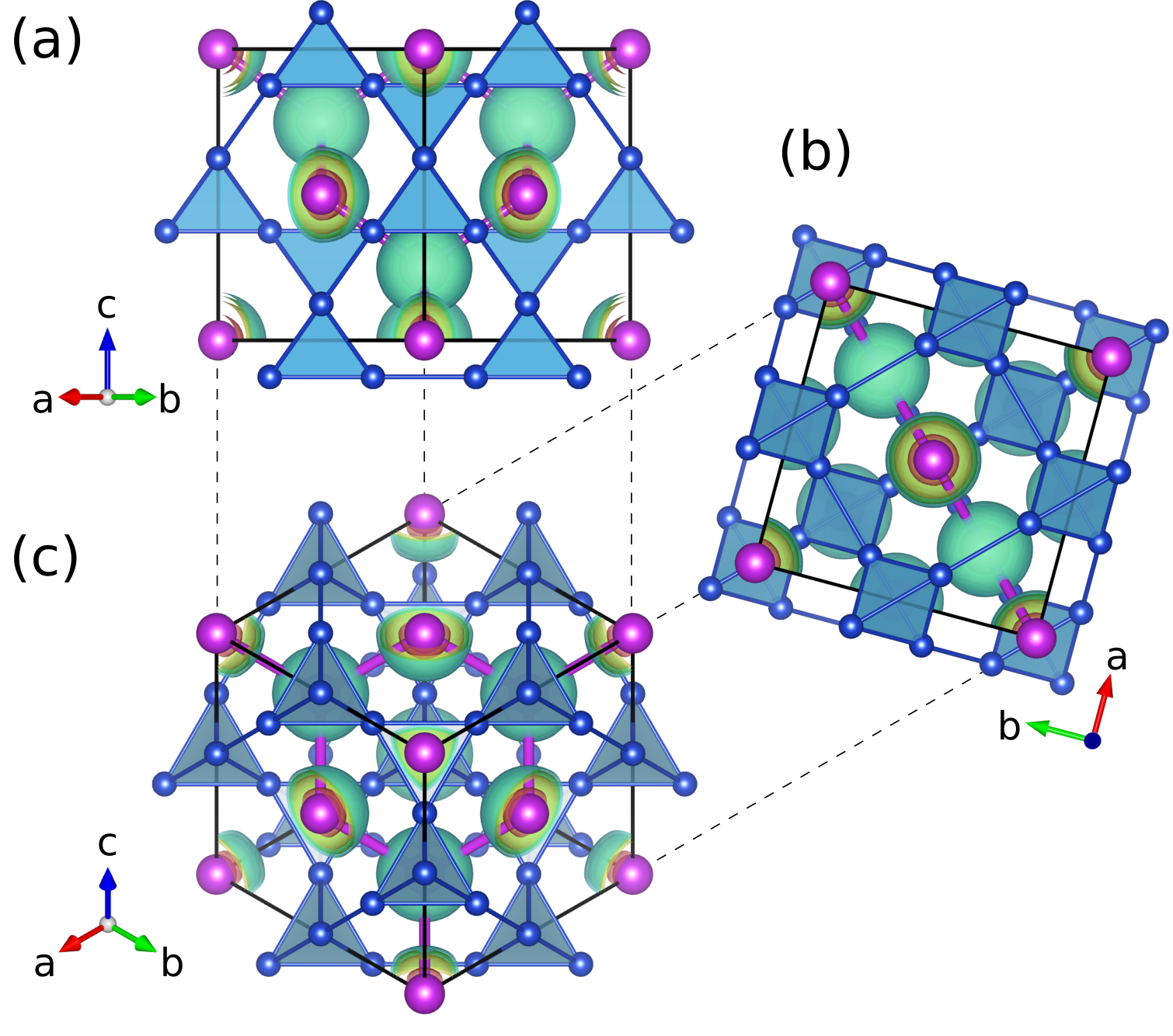}
\caption{\label{fig:structure} The crystal structure of cubic\ce{Cu2Bi} from three different perspectives is shown in the panels (a), (b) and (c). Purple (large) and blue (small) spheres denote the Bi and Cu atoms, respectively.  The isosurfaces of the ELF are shown at values of 0.9 (red), 0.8 (yellow), 0.7 (green), and 0.6 (blue).}
\end{figure}

Although at first glance the channels and voids in  \ce{Cu11Bi7} and CuBi appear to be empty, they serve a specific purpose in these materials, namely to provide room for hosting the stereochemically active bismuth lone pairs~\cite{Clarke2016,clarke_creating_2017}. The formation of such porous structures at high pressures is somewhat surprising, since the thermodynamic stability of a solid state compound is governed by the Gibbs free energy, $G=E+pV-TS$, which tends to favor dense structures as the $pV$ term becomes increasingly dominant at higher pressures. Many examples are known where pressure leads to a collapse of low-dimensional structures towards a polymorph with higher packing density, and the transition from graphite to cubic diamond is only one example~\cite{naka_direct_1976}. Hence, we suspect that also in the Cu--Bi system a high-density phase awaits discovery that becomes accessible once a  sufficiently high pressure is applied.

To identify these potential high-density phases, we carried out a search by employing the Minima Hopping structure prediction method (MHM) at 10 and 150~GPa with the \texttt{Minhocao} package~\cite{goedecker2004,amsler2010}, which implements a highly reliable algorithm to explore the low enthalpy phases of a compound at a specific pressure solely given the chemical composition~\cite{amsler_crystal_2012,flores-livas_high-pressure_2012,amsler_novel_2012,botti_carbon_2013,huan_thermodynamic_2013}. Within this method, the low lying part of the enthalpy landscape is efficiently sampled by performing consecutive, short molecular dynamics (MD) escape steps to overcome enthalpy barriers, followed by local geometry optimizations. The Bell-Evans-Polanyi principle is exploited by aligning the initial MD velocities along soft mode directions in order to accelerate the search~\cite{roy_bell-evans-polanyi_2008,sicher_efficient_2011}. In the current study, the enthalpy landscape was modeled at the density functional theory (DFT) level, using the projector augmented wave~(PAW) formalism~\cite{PAW-Blochl-1994} as implemented in the \texttt{VASP}~\cite{VASP-Kresse-1995,VASP-Kresse-1996,VASP-Kresse-1999} package together with the Perdew-Burke-Ernzerhof (PBE) approximation~\cite{Perdew-PBE-1996} to the exchange correlation potential. The most promising candidate structures were refined by performing relaxations in intervals of 10~GPa with a plane-wave cutoff energy of 400~eV and a sufficiently dense k-point mesh to ensure a convergence of the total energy to within 2~meV/atom

\begin{figure}[tb]
\includegraphics[width=1\columnwidth]{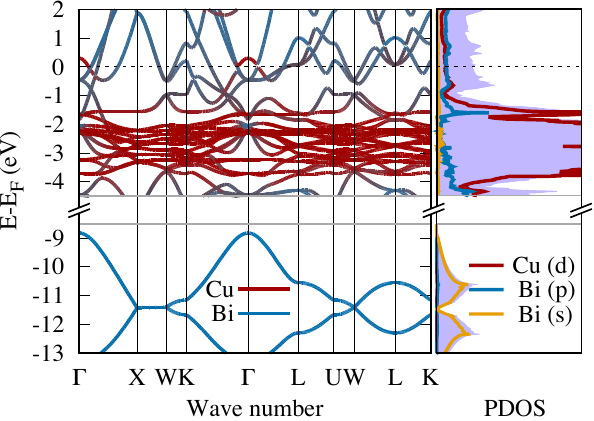}
\caption{\label{fig:bands} The electronic band structure in the irreducible Brillouin zone is shown, color coded according to the projection on the Cu and Bi atoms in red and blue, respectively. In the right panel, the total density of states is shown as the shaded area, and the contributions of the Cu $d$-states, Bi $p$-states and Bi $s$-states are indicated by the red, blue and orange lines.}
\end{figure}

We performed structural searches at variable composition for systems with up to 20 atoms/cell. As intuitively expected, we found novel phases with higher packing densities that become thermodynamically stable at pressures readily accessible with current high-pressure techniques using diamond anvil cells (DAC), i.e. below 100~GPa. A tetragonal \ce{Cu2Bi} compound with $I4/mmm$ symmetry of \ce{La2Sb} type, isostructural to \ce{Ti2Bi}~\cite{auer-welsbach_untersuchung_1958}, touches the convex hull of stability at 51~GPa. Its pressure range of stability is however very small and reaches merely up to 59~GPa, above which another \ce{Cu2Bi} phase in the cubic Laves structure (Strukturbericht designation: C15) with \textit{Fd-3m} symmetry becomes stable. Since an ordered representation is used to model the disordered \ce{Cu11Bi7} phase~\cite{Clarke2016}, we believe that our calculations overestimate its formation enthalpy, such that the true stability range of the tetragonal \ce{Cu2Bi} is even smaller and therefore this phase might not be synthesizeable at all in practice. On the other hand, the cubic \ce{Cu2Bi} phase has an extended range of stability up to several hundred GPa, rendering it the most promising candidate structure for \ce{Cu2Bi}. We also investigated the two other Laves phases, the hexagonal C14 and C36 structures, which were both found to have higher enthalpies than C15. The structure of the cubic \ce{Cu2Bi} is shown in Fig.~\ref{fig:structure} from three different perspectives. A single Cu atom occupies the $16c$ Wyckoff site at $(0,0,0)$, while a Bi atom occupies the $8 b$ site at $(0.375,0.375, 0.375)$. The lattice constant is $6.77$~\AA\, at 60~GPa, and $7.52$~\AA\, at ambient pressure, respectively. This crystal structure, which was also observed in the isoelectronic \ce{Au2Bi} compound~\cite{jurriaanse_crystal_2015}, can be interpreted as two interpenetrating sub-lattices of Cu and Bi, where the Bi atoms are arranged in a face centered cubic diamond structure, whereas closely packed Cu tetrahedra occupy the tetrahedral ($T_d$) interstitial sites of the Bi sublattice. The centers of these tetrahedra themselves compose the second diamond sublattice, which are shown as blue polyhedra in Fig.~\ref{fig:structure}.

This picture of interpenetrating diamond lattices is merely an  interpretation of the structure, since there are no  covalent bonds between the Bi atoms as one would expect from $sp^3$ materials like carbon diamond, and the Bi--Bi distance is as large as 3.254~\AA\, at 0~GPa which is significantly higher than the covalent bond length of a single Bi--Bi bond of 3.02~\AA. The electronic band structure in Fig.~\ref{fig:bands} shows that the cubic \ce{Cu2Bi} is metallic, and like in the two experimentally observed Cu--Bi compounds \ce{Cu11Bi7} and CuBi, the bands at the Fermi level of cubic \ce{Cu2Bi}  are dominated by the Cu $d$ and Bi $p$-states, as illustrated by the partial density of states (PDOS) in the right panel. The  Bi $6s$ lone pairs are located deep below the Fermi level, as indicated by the orange lines at the bottom of the PDOS.

Laves phases are known for their high packing efficiencies, and according to our calculations cubic \ce{Cu2Bi} has a density of $\rho=10.52$~g/cm$^{3}$ if recovered to ambient pressures. This value is considerably higher than the computed densities of CuBi ($\rho=10.03$~g/cm$^{3}$) and \ce{Cu11Bi7} ($\rho= 10.28$~g/cm$^{3}$), but also than the decomposition product 2Cu+Bi ($\rho=9.19$~g/cm$^{3}$). The dense packing in cubic \ce{Cu2Bi} suggests that its bonding properties differ strongly from the two previously reported Cu--Bi compounds \ce{Cu11Bi7} and CuBi. Indeed, the isosurfaces of the electron localization function (ELF) of cubic \ce{Cu2Bi} in Fig.~\ref{fig:structure}  reveal that the Bi 6$s^2$ electrons are localized with spherical symmetry  around the bismuth nuclei. Hence, the Bi lone pairs are stereochemically inactive, in strong contrast to both \ce{Cu11Bi7} and CuBi.

\begin{figure}[tb]
\subfloat[Stability range\label{sfig:stability}]{%
\includegraphics[width=1\columnwidth]{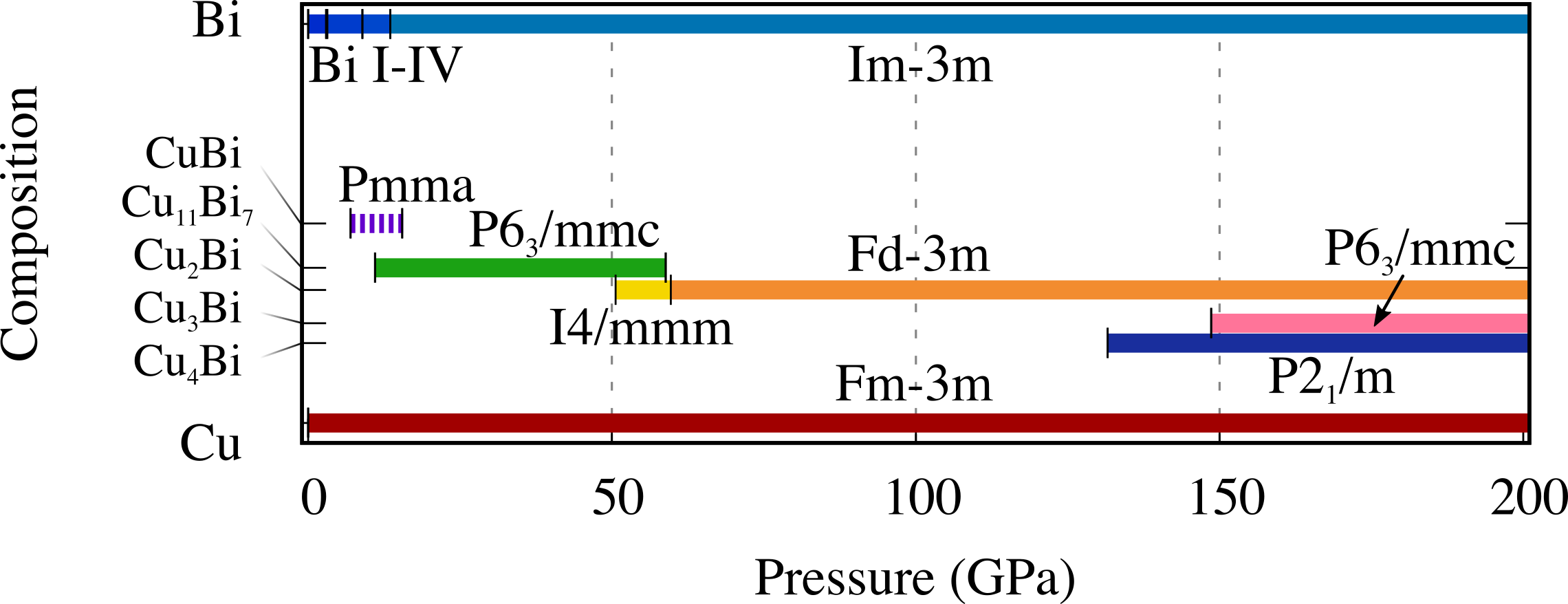}
}\\
\subfloat[Convex hull\label{sfig:hull}]{%
\includegraphics[width=1\columnwidth]{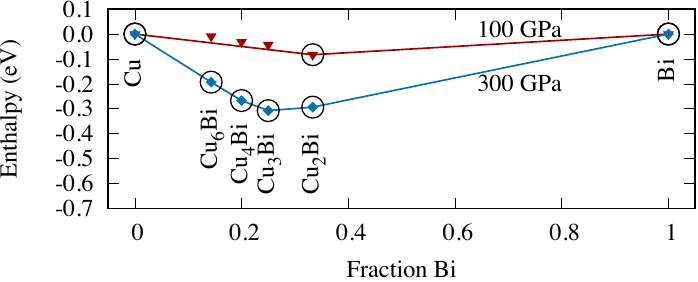}
}
\caption{\label{fig:stability} (a) The stability range of the  different Cu--Bi polymorphs are shown as a function of pressure. The dashed line representing the CuBi phase indicates that it is only stable at elevated temperatures. (b) The convex hull of stability at 100~GPa (red) and at 300~GPa (blue). Thermodynamically stable phases are denoted by black circles.}
\end{figure}

From the energetic point of view, cubic \ce{Cu2Bi} is metastable with respect to elemental decomposition by about 160~meV/atom at ambient conditions. Although this value is considerably higher than for \ce{Cu11Bi7} (55~meV/atom) or CuBi (48~meV/atom), it is within the energy window of observed metastable materials~\cite{sun_thermodynamic_2016}. In comparison, the recently predicted high-pressure  phase \ce{FeBi2} (transition pressure 36~GPa) in a similar chemical system has a formation energy of above 240~meV/atom at 0~GPa~\cite{amsler_prediction_2017}, but was nevertheless synthesized and quenched to as low as 3~GPa~\cite{walsh_discovery_2016}. Since the MHM explores the energy landscape using physical MD moves, we additionally performed six short MHM simulation at ambient pressure to assess the kinetic stability of cubic \ce{Cu2Bi}, starting from the C15 structure in a 12 atom cell. The lowest energy structure found within the first successful MD escape trial is roughly 30~meV/atom higher in energy than the C15 phase (found by 5 out of the 6 runs), requiring a kinetic energy corresponding to around 1200~K. Although this value doesn't directly correspond to a physical temperature, it gives a rough estimate of the upper bound for the transition barrier. The preferred escape towards a higher energy state indicates that there is no direct downhill path to a lower energy structure, and therefore that the C15 structure is in a well of a funnel surrounded by high barriers~\cite{wales_energy_2004}. Hence, we expect that cubic \ce{Cu2Bi} can be quenched and recovered to ambient pressure as a third metastable Cu--Bi phase.

Fig.~\ref{sfig:stability} shows the pressure range of stability of all Cu--Bi phases known to date, together with the two \ce{Cu2Bi} structures, as a function of pressure, derived from the formation enthalpy at zero temperature.  Note that the bar representing CuBi is drawn as dashed line to indicate that it is not a stable phase at 0~K, but only becomes thermodynamically accessible at elevated temperatures as we recently demonstrated in Ref.~\onlinecite{clarke_creating_2017}. A striking feature of Fig.~\ref{sfig:stability} is that, already at moderate pressures up to 100 GPa, phases with a higher Cu content are favored as the pressure increases. This trend can be explained by a simple argument based on the large difference in atomic radii of Cu and Bi: a preferred high packing density at extreme pressures is only possible when the fraction of the smaller Cu atoms increases to fill the gaps between the large Bi atoms.

This trend carries on for the phase space at pressures exceeding 100~GPa, where we discovered at least three additional Cu--Bi compounds that become thermodynamically stable  with even larger Cu content than $\frac{2}{3}$, as shown in Fig.~\ref{sfig:stability}. First, a \ce{Cu4Bi} phase touches the convex hull of stability at a pressure of about 130~GPa, which crystallizes in a monoclinic \textit{P2$_1$/m} structure. At 148~GPa, a hexagonal structure with \textit{P6$_3$/mmc} symmetry becomes stable at the \ce{Cu3Bi} composition. Finally, we find a \ce{Cu6Bi} phase in a  \textit{P2$_1$/m} structure at pressures above 267~GPa, as shown by the  blue convex hull plot in Fig.~\ref{sfig:hull}.  The structural features of those three phases strongly differ from CuBi, \ce{Cu11Bi7} and both \ce{Cu2Bi} phases. They all share the same motif of Bi atoms coordinated to 12 Cu atoms, which form the corners of a cuboctahedron. In fact, all structures are identical to those previously reported in the \ce{XeNi_n} and \ce{XeFe_n} systems~\cite{zhu_reactions_2014}, where the individual cuboctahedra are linked together in different geometries. The detailed structural data are given in the supplemental materials.

\begin{figure}[tb]
\includegraphics[width=1\columnwidth]{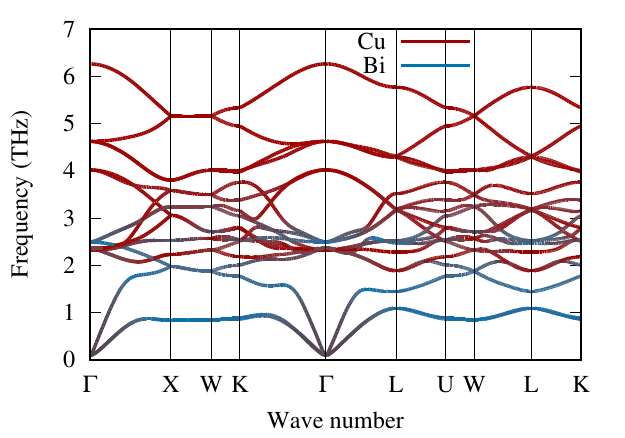}
\caption{\label{fig:phonon} The phonon band structure is shown along a path in the  Brillouin zone. The color coding represents the factional mode contribution of the Cu and Bi atoms to the phonon eigenmodes in red and blue, respectively.}
\end{figure}

Since cubic \ce{Cu2Bi} is the phase that becomes stable at the lowest pressure besides \ce{Cu11Bi7} and CuBi with a wide pressure range of stability, we will from hereon focus solely on a detailed characterization of this phase. To assess if cubic \ce{Cu2Bi} is dynamically stable at ambient conditions, i.e. if it corresponds to a local minimum on the energy landscape, we computed its phonon dispersion in the whole Brillouin zone using linear response calculations as implemented in the \texttt{Quantum Espresso} package~\cite{espresso}.  Norm conserving pseudopotentials were used with a plane-wave cutoff energy of 150~Ry, and the force constants were evaluated on a $q$-grid of $4\times4\times4$. Fig.~\ref{fig:phonon} shows that all phonon frequencies are real, indicating that the structure is indeed dynamically stable. The bands are colored according to the contributions of the Cu and Bi atoms to the phonon eigenmodes at each wave vector. Similar to \ce{Cu11Bi7} and CuBi, the low energy phonons are dominated by the vibration of the heavy Bi atoms, while the high-frequency modes stem mainly from the Cu atoms. 

\begin{figure}[tb]
\includegraphics[width=1\columnwidth]{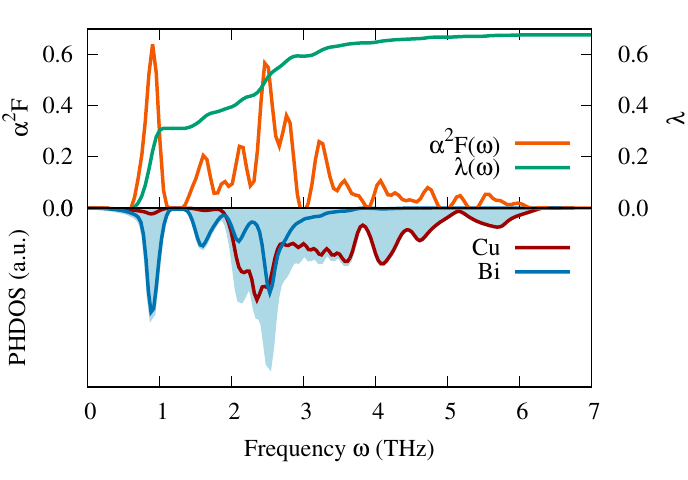}
\caption{\label{fig:elphon} The electron-phonon coupling properties are shown in the top panel as a function of the phonon frequency $\omega$, where $\alpha^2F(\omega)$ denotes the Eliashberg spectral function, while $\lambda(\omega)$ is the electron-phonon coupling strength. In the bottom panel the total phonon density of states (PHDOS) is shown with the shaded area, whereas the partial density of states of the Cu and the Bi atoms are shown in red and blue, respectively.}
\end{figure}

We find a remarkable feature in the phonon band structure, where  the lowest energy, doubly degenerate transversal acoustic branch is highly localized and does not cross any optical bands. Such vibrations are often referred to as rattling modes and can significantly affect materials properties~\cite{he_ultralow_2016}. In clathrate materials such as \ce{Ba8Si46}, where the guest Ba atoms reside in the cages of the Si host structure, such rattling modes promote the  electron-phonon coupling~\cite{tse_electronic_2005} and lead to its high superconducting temperature of $T_c=8$~K~\cite{yamanaka_high-pressure_2000,tanigaki_mechanism_2003}. In a similar manner, the low energy Bi vibrations contribute to the electron-phonon coupling in cubic \ce{Cu2Bi}, as we will show below. 

The superconducting transition temperature was computed within the Allan-Dynes modified McMillan's approximation of the Eliashberg equation~\cite{Allen_1975}, using a Coulomb pseudopotential value of  $\mu^{*}=0.13$ and a dense $k$-mesh of $24\times24\times24$. The resulting overall electron-phonon coupling parameter  $\lambda=0.68$ leads to a moderate $T_c$ of 2.0~K at 0~GPa, which is slightly higher than in \ce{Cu11Bi7} and CuBi. We can relate this difference in $T_c$ to the different bonding behavior. As shown in Fig.~\ref{fig:elphon}, a large contribution to the electron-phonon coupling constant $\lambda$ stems from a peak in the Eliashberg spectral function $\alpha^2F$, located exactly at the frequency of the Bi rattling mode as shown in the lower panel. Although a similar electron-phonon coupling of Bi vibrations is observed both in \ce{Cu11Bi7}~\cite{Clarke2016} and CuBi~\cite{clarke_creating_2017}, there is an additional contribution to $\lambda$ from phonons between 2 and 4~THz in cubic \ce{Cu2Bi}, giving rise to the higher value in $T_c$.

In summary,  we predict from \textit{ab initio} calculations a set of high-pressure intermetallic compounds in the Cu--Bi system besides the previously synthesized phases, CuBi and \ce{Cu11Bi7}. The compound which becomes stable at the lowest pressures is tetragonal \ce{Cu2Bi}, but the most promising candidate is cubic \ce{Cu2Bi} which crystallizes in a Laves structure above 59~GPa. In strong contrast to the recently reported \ce{Cu11Bi7} and CuBi, which both contain voids to host the stereochemically active Bi lone pairs, the structure of cubic \ce{Cu2Bi} allows a dense packing of the constituent atoms. In agreement with the common perception that materials with voids and low-density structures become increasingly unfavorable at high pressures, cubic \ce{Cu2Bi} is predicted to be thermodynamically accessible at pressures exceeding 59~GPa due to its high volumetric density. Additionally, cubic \ce{Cu2Bi} is dynamically stable and we provide evidence that the system is trapped at the bottom of a funnel on the energy landscape, indicating that it can be recovered to ambient pressure. A rattling mode in the phonon band structure of cubic \ce{Cu2Bi} couples strongly to the electrons, leading to a conventional superconducting transition temperature of $T_c=2.0$~K exceeding the values in \ce{Cu11Bi7} and CuBi. At  higher pressures above 100~GPa, three additional phases with an even higher copper content are predicted to become thermodynamically stable, namely \ce{Cu3Bi}, \ce{Cu4Bi}, and \ce{Cu6Bi}. These phases share the same structural motif of interlinked cuboctahedra with 12-coordinated Bi atoms. Overall, the current study contributes to the recent efforts in exploring the phase space of the ambient-immiscible Cu--Bi system, which still bears many potential high-pressure phases awaiting discovery.

\section{Acknowledgments}
M.A. acknowledges support from the Novartis Universit{\"a}t Basel Excellence Scholarship for Life Sciences and the Swiss National Science Foundation (P300P2-158407). C.W. acknowledges support by the U.S. Department of Energy, Office of Science, Basic Energy Sciences, under Grant No. DE-FG02-07ER46433. Computing resources from the following centers are gratefully acknowledged: the Swiss National Supercomputing Center in Lugano (project s700), the Extreme Science and Engineering Discovery Environment (XSEDE) (which is supported by National Science Foundation grant number OCI-1053575), the Bridges system at the Pittsburgh Supercomputing Center (PSC) (which is supported by NSF award number ACI-1445606), the Quest high performance computing facility at Northwestern University, and the National Energy Research Scientific Computing Center (DOE: DE-AC02-05CH11231). 



\begin{thebibliography}{52}%
\makeatletter
\providecommand \@ifxundefined [1]{%
 \@ifx{#1\undefined}
}%
\providecommand \@ifnum [1]{%
 \ifnum #1\expandafter \@firstoftwo
 \else \expandafter \@secondoftwo
 \fi
}%
\providecommand \@ifx [1]{%
 \ifx #1\expandafter \@firstoftwo
 \else \expandafter \@secondoftwo
 \fi
}%
\providecommand \natexlab [1]{#1}%
\providecommand \enquote  [1]{``#1''}%
\providecommand \bibnamefont  [1]{#1}%
\providecommand \bibfnamefont [1]{#1}%
\providecommand \citenamefont [1]{#1}%
\providecommand \href@noop [0]{\@secondoftwo}%
\providecommand \href [0]{\begingroup \@sanitize@url \@href}%
\providecommand \@href[1]{\@@startlink{#1}\@@href}%
\providecommand \@@href[1]{\endgroup#1\@@endlink}%
\providecommand \@sanitize@url [0]{\catcode `\\12\catcode `\$12\catcode
  `\&12\catcode `\#12\catcode `\^12\catcode `\_12\catcode `\%12\relax}%
\providecommand \@@startlink[1]{}%
\providecommand \@@endlink[0]{}%
\providecommand \url  [0]{\begingroup\@sanitize@url \@url }%
\providecommand \@url [1]{\endgroup\@href {#1}{\urlprefix }}%
\providecommand \urlprefix  [0]{URL }%
\providecommand \Eprint [0]{\href }%
\providecommand \doibase [0]{http://dx.doi.org/}%
\providecommand \selectlanguage [0]{\@gobble}%
\providecommand \bibinfo  [0]{\@secondoftwo}%
\providecommand \bibfield  [0]{\@secondoftwo}%
\providecommand \translation [1]{[#1]}%
\providecommand \BibitemOpen [0]{}%
\providecommand \bibitemStop [0]{}%
\providecommand \bibitemNoStop [0]{.\EOS\space}%
\providecommand \EOS [0]{\spacefactor3000\relax}%
\providecommand \BibitemShut  [1]{\csname bibitem#1\endcsname}%
\let\auto@bib@innerbib\@empty
\bibitem [{\citenamefont {Nie}(2012)}]{nie_precipitation_2012}%
  \BibitemOpen
  \bibfield  {author} {\bibinfo {author} {\bibfnamefont {J.-F.}\ \bibnamefont
  {Nie}},\ }\href {\doibase 10.1007/s11661-012-1217-2} {\bibfield  {journal}
  {\bibinfo  {journal} {Metall and Mat Trans A}\ }\textbf {\bibinfo {volume}
  {43}},\ \bibinfo {pages} {3891} (\bibinfo {year} {2012})}\BibitemShut
  {NoStop}%
\bibitem [{\citenamefont {Wang}\ and\ \citenamefont
  {Starink}(2005)}]{wang_precipitates_2005}%
  \BibitemOpen
  \bibfield  {author} {\bibinfo {author} {\bibfnamefont {S.~C.}\ \bibnamefont
  {Wang}}\ and\ \bibinfo {author} {\bibfnamefont {M.~J.}\ \bibnamefont
  {Starink}},\ }\href {\doibase Wang, S.C. and Starink, M.J. (2005)
  Precipitates and intermetallic phases in precipitation hardening
  Al–Cu–Mg–(Li) based alloys International Materials Review, 50, (4), pp.
  193-215. (doi:10.1179/174328005X14357
  <http://dx.doi.org/10.1179/174328005X14357>).} {\bibfield  {journal}
  {\bibinfo  {journal} {International Materials Review}\ }\textbf {\bibinfo
  {volume} {50}},\ \bibinfo {pages} {193} (\bibinfo {year} {2005})}\BibitemShut
  {NoStop}%
\bibitem [{\citenamefont {Fidler}\ \emph {et~al.}(2007)\citenamefont {Fidler},
  \citenamefont {Suess},\ and\ \citenamefont
  {Schrefl}}]{fidler_rare-earth_2007}%
  \BibitemOpen
  \bibfield  {author} {\bibinfo {author} {\bibfnamefont {J.}~\bibnamefont
  {Fidler}}, \bibinfo {author} {\bibfnamefont {D.}~\bibnamefont {Suess}}, \
  and\ \bibinfo {author} {\bibfnamefont {T.}~\bibnamefont {Schrefl}},\ }in\
  \href
  {http://onlinelibrary.wiley.com/doi/10.1002/9780470022184.hmm404/abstract}
  {\emph {\bibinfo {booktitle} {Handbook of Magnetism and Advanced Magnetic
  Materials}}}\ (\bibinfo  {publisher} {John Wiley \& Sons, Ltd},\ \bibinfo
  {year} {2007})\BibitemShut {NoStop}%
\bibitem [{\citenamefont {Croat}\ \emph {et~al.}(1984)\citenamefont {Croat},
  \citenamefont {Herbst}, \citenamefont {Lee},\ and\ \citenamefont
  {Pinkerton}}]{Croat1984}%
  \BibitemOpen
  \bibfield  {author} {\bibinfo {author} {\bibfnamefont {J.~J.}\ \bibnamefont
  {Croat}}, \bibinfo {author} {\bibfnamefont {J.~F.}\ \bibnamefont {Herbst}},
  \bibinfo {author} {\bibfnamefont {R.~W.}\ \bibnamefont {Lee}}, \ and\
  \bibinfo {author} {\bibfnamefont {F.~E.}\ \bibnamefont {Pinkerton}},\ }\href
  {\doibase 10.1063/1.333571} {\bibfield  {journal} {\bibinfo  {journal} {J.
  Appl. Phys.}\ }\textbf {\bibinfo {volume} {55}},\ \bibinfo {pages} {2078}
  (\bibinfo {year} {1984})}\BibitemShut {NoStop}%
\bibitem [{\citenamefont {Sagawa}\ \emph {et~al.}(1984)\citenamefont {Sagawa},
  \citenamefont {Fujimura}, \citenamefont {Togawa}, \citenamefont {Yamamoto},\
  and\ \citenamefont {Matsuura}}]{Sagawa1984}%
  \BibitemOpen
  \bibfield  {author} {\bibinfo {author} {\bibfnamefont {M.}~\bibnamefont
  {Sagawa}}, \bibinfo {author} {\bibfnamefont {S.}~\bibnamefont {Fujimura}},
  \bibinfo {author} {\bibfnamefont {N.}~\bibnamefont {Togawa}}, \bibinfo
  {author} {\bibfnamefont {H.}~\bibnamefont {Yamamoto}}, \ and\ \bibinfo
  {author} {\bibfnamefont {Y.}~\bibnamefont {Matsuura}},\ }\href {\doibase
  10.1063/1.333572} {\bibfield  {journal} {\bibinfo  {journal} {J. Appl.
  Phys.}\ }\textbf {\bibinfo {volume} {55}},\ \bibinfo {pages} {2083} (\bibinfo
  {year} {1984})}\BibitemShut {NoStop}%
\bibitem [{\citenamefont {Benz}\ and\ \citenamefont {Martin}(1970)}]{Benz1970}%
  \BibitemOpen
  \bibfield  {author} {\bibinfo {author} {\bibfnamefont {M.~G.}\ \bibnamefont
  {Benz}}\ and\ \bibinfo {author} {\bibfnamefont {D.~L.}\ \bibnamefont
  {Martin}},\ }\href {\doibase 10.1063/1.1653354} {\bibfield  {journal}
  {\bibinfo  {journal} {Appl. Phys. Lett.}\ }\textbf {\bibinfo {volume} {17}},\
  \bibinfo {pages} {176} (\bibinfo {year} {1970})}\BibitemShut {NoStop}%
\bibitem [{\citenamefont {Amsler}\ \emph
  {et~al.}(2017{\natexlab{a}})\citenamefont {Amsler}, \citenamefont {Naghavi},\
  and\ \citenamefont {Wolverton}}]{amsler_prediction_2017}%
  \BibitemOpen
  \bibfield  {author} {\bibinfo {author} {\bibfnamefont {M.}~\bibnamefont
  {Amsler}}, \bibinfo {author} {\bibfnamefont {S.~S.}\ \bibnamefont {Naghavi}},
  \ and\ \bibinfo {author} {\bibfnamefont {C.}~\bibnamefont {Wolverton}},\
  }\href {\doibase 10.1039/C6SC04683E} {\bibfield  {journal} {\bibinfo
  {journal} {Chem. Sci.}\ }\textbf {\bibinfo {volume} {8}},\ \bibinfo {pages}
  {2226} (\bibinfo {year} {2017}{\natexlab{a}})}\BibitemShut {NoStop}%
\bibitem [{\citenamefont {Walsh}\ \emph {et~al.}(2016)\citenamefont {Walsh},
  \citenamefont {Clarke}, \citenamefont {Meng}, \citenamefont {Jacobsen},\ and\
  \citenamefont {Freedman}}]{walsh_discovery_2016}%
  \BibitemOpen
  \bibfield  {author} {\bibinfo {author} {\bibfnamefont {J.~P.~S.}\
  \bibnamefont {Walsh}}, \bibinfo {author} {\bibfnamefont {S.~M.}\ \bibnamefont
  {Clarke}}, \bibinfo {author} {\bibfnamefont {Y.}~\bibnamefont {Meng}},
  \bibinfo {author} {\bibfnamefont {S.~D.}\ \bibnamefont {Jacobsen}}, \ and\
  \bibinfo {author} {\bibfnamefont {D.~E.}\ \bibnamefont {Freedman}},\ }\href
  {\doibase 10.1021/acscentsci.6b00287} {\bibfield  {journal} {\bibinfo
  {journal} {ACS Cent. Sci.}\ }\textbf {\bibinfo {volume} {2}},\ \bibinfo
  {pages} {867} (\bibinfo {year} {2016})}\BibitemShut {NoStop}%
\bibitem [{\citenamefont {Sturza}\ \emph {et~al.}(2014)\citenamefont {Sturza},
  \citenamefont {Han}, \citenamefont {Malliakas}, \citenamefont {Chung},
  \citenamefont {Claus},\ and\ \citenamefont
  {Kanatzidis}}]{sturza_superconductivity_2014}%
  \BibitemOpen
  \bibfield  {author} {\bibinfo {author} {\bibfnamefont {M.}~\bibnamefont
  {Sturza}}, \bibinfo {author} {\bibfnamefont {F.}~\bibnamefont {Han}},
  \bibinfo {author} {\bibfnamefont {C.~D.}\ \bibnamefont {Malliakas}}, \bibinfo
  {author} {\bibfnamefont {D.~Y.}\ \bibnamefont {Chung}}, \bibinfo {author}
  {\bibfnamefont {H.}~\bibnamefont {Claus}}, \ and\ \bibinfo {author}
  {\bibfnamefont {M.~G.}\ \bibnamefont {Kanatzidis}},\ }\href {\doibase
  10.1103/PhysRevB.89.054512} {\bibfield  {journal} {\bibinfo  {journal} {Phys.
  Rev. B}\ }\textbf {\bibinfo {volume} {89}},\ \bibinfo {pages} {054512}
  (\bibinfo {year} {2014})}\BibitemShut {NoStop}%
\bibitem [{\citenamefont {Dong}\ and\ \citenamefont
  {Fan}(2015)}]{dong_rich_2015}%
  \BibitemOpen
  \bibfield  {author} {\bibinfo {author} {\bibfnamefont {X.}~\bibnamefont
  {Dong}}\ and\ \bibinfo {author} {\bibfnamefont {C.}~\bibnamefont {Fan}},\
  }\href {\doibase 10.1038/srep09326} {\bibfield  {journal} {\bibinfo
  {journal} {Sci. Rep.}\ }\textbf {\bibinfo {volume} {5}},\ \bibinfo {pages}
  {9326} (\bibinfo {year} {2015})}\BibitemShut {NoStop}%
\bibitem [{\citenamefont {Haegg}\ and\ \citenamefont
  {Funke}(1929)}]{Haegg1929}%
  \BibitemOpen
  \bibfield  {author} {\bibinfo {author} {\bibfnamefont {G.}~\bibnamefont
  {Haegg}}\ and\ \bibinfo {author} {\bibfnamefont {G.}~\bibnamefont {Funke}},\
  }\href@noop {} {\bibfield  {journal} {\bibinfo  {journal} {Z. Phys. Chem.,
  Abt. B}\ }\textbf {\bibinfo {volume} {6}},\ \bibinfo {pages} {272} (\bibinfo
  {year} {1929})}\BibitemShut {NoStop}%
\bibitem [{\citenamefont {Glagoleva}\ and\ \citenamefont
  {Zhdanov}(1954)}]{Glagoleva1954}%
  \BibitemOpen
  \bibfield  {author} {\bibinfo {author} {\bibfnamefont {V.~P.}\ \bibnamefont
  {Glagoleva}}\ and\ \bibinfo {author} {\bibfnamefont {G.~S.}\ \bibnamefont
  {Zhdanov}},\ }\href@noop {} {\bibfield  {journal} {\bibinfo  {journal} {Zh.
  Eksp. Teor. Fiz.}\ }\textbf {\bibinfo {volume} {26}},\ \bibinfo {pages} {337}
  (\bibinfo {year} {1954})}\BibitemShut {NoStop}%
\bibitem [{\citenamefont {Ruck}\ and\ \citenamefont
  {S\"{o}hnel}(2006)}]{Ruck2006}%
  \BibitemOpen
  \bibfield  {author} {\bibinfo {author} {\bibfnamefont {M.}~\bibnamefont
  {Ruck}}\ and\ \bibinfo {author} {\bibfnamefont {T.}~\bibnamefont
  {S\"{o}hnel}},\ }\href {\doibase 10.1515/znb-2006-0703} {\bibfield  {journal}
  {\bibinfo  {journal} {Z. Naturforsch.}\ }\textbf {\bibinfo {volume} {61b}},\
  \bibinfo {pages} {785} (\bibinfo {year} {2006})}\BibitemShut {NoStop}%
\bibitem [{\citenamefont {Matthias}\ \emph {et~al.}(1966)\citenamefont
  {Matthias}, \citenamefont {Jayaraman}, \citenamefont {Geballe}, \citenamefont
  {Andres},\ and\ \citenamefont {Corenzwit}}]{Matthias1961}%
  \BibitemOpen
  \bibfield  {author} {\bibinfo {author} {\bibfnamefont {B.~T.}\ \bibnamefont
  {Matthias}}, \bibinfo {author} {\bibfnamefont {A.}~\bibnamefont {Jayaraman}},
  \bibinfo {author} {\bibfnamefont {T.~H.}\ \bibnamefont {Geballe}}, \bibinfo
  {author} {\bibfnamefont {K.}~\bibnamefont {Andres}}, \ and\ \bibinfo {author}
  {\bibfnamefont {E.}~\bibnamefont {Corenzwit}},\ }\href {\doibase
  10.1103/PhysRevLett.17.640} {\bibfield  {journal} {\bibinfo  {journal} {Phys.
  Rev. Lett.}\ }\textbf {\bibinfo {volume} {17}},\ \bibinfo {pages} {640}
  (\bibinfo {year} {1966})}\BibitemShut {NoStop}%
\bibitem [{\citenamefont {Schwarz}\ \emph {et~al.}(2013)\citenamefont
  {Schwarz}, \citenamefont {Tenc{\'{e}}}, \citenamefont {Janson}, \citenamefont
  {Koz}, \citenamefont {Krellner}, \citenamefont {Burkhardt}, \citenamefont
  {Rosner}, \citenamefont {Steglich},\ and\ \citenamefont
  {Grin}}]{Schwarz2013}%
  \BibitemOpen
  \bibfield  {author} {\bibinfo {author} {\bibfnamefont {U.}~\bibnamefont
  {Schwarz}}, \bibinfo {author} {\bibfnamefont {S.}~\bibnamefont
  {Tenc{\'{e}}}}, \bibinfo {author} {\bibfnamefont {O.}~\bibnamefont {Janson}},
  \bibinfo {author} {\bibfnamefont {C.}~\bibnamefont {Koz}}, \bibinfo {author}
  {\bibfnamefont {C.}~\bibnamefont {Krellner}}, \bibinfo {author}
  {\bibfnamefont {U.}~\bibnamefont {Burkhardt}}, \bibinfo {author}
  {\bibfnamefont {H.}~\bibnamefont {Rosner}}, \bibinfo {author} {\bibfnamefont
  {F.}~\bibnamefont {Steglich}}, \ and\ \bibinfo {author} {\bibfnamefont
  {Y.}~\bibnamefont {Grin}},\ }\href {\doibase 10.1002/anie.201302397}
  {\bibfield  {journal} {\bibinfo  {journal} {Angew. Chemie Int. Ed.}\ }\textbf
  {\bibinfo {volume} {52}},\ \bibinfo {pages} {9853} (\bibinfo {year}
  {2013})}\BibitemShut {NoStop}%
\bibitem [{\citenamefont {Tenc{\'{e}}}\ \emph {et~al.}(2014)\citenamefont
  {Tenc{\'{e}}}, \citenamefont {Janson}, \citenamefont {Krellner},
  \citenamefont {Rosner}, \citenamefont {Schwarz}, \citenamefont {Grin},\ and\
  \citenamefont {Steglich}}]{Tence2014}%
  \BibitemOpen
  \bibfield  {author} {\bibinfo {author} {\bibfnamefont {S.}~\bibnamefont
  {Tenc{\'{e}}}}, \bibinfo {author} {\bibfnamefont {O.}~\bibnamefont {Janson}},
  \bibinfo {author} {\bibfnamefont {C.}~\bibnamefont {Krellner}}, \bibinfo
  {author} {\bibfnamefont {H.}~\bibnamefont {Rosner}}, \bibinfo {author}
  {\bibfnamefont {U.}~\bibnamefont {Schwarz}}, \bibinfo {author} {\bibfnamefont
  {Y.}~\bibnamefont {Grin}}, \ and\ \bibinfo {author} {\bibfnamefont
  {F.}~\bibnamefont {Steglich}},\ }\href {\doibase
  10.1088/0953-8984/26/39/395701} {\bibfield  {journal} {\bibinfo  {journal}
  {J. Phys. Condens. Matter}\ }\textbf {\bibinfo {volume} {26}},\ \bibinfo
  {pages} {395701} (\bibinfo {year} {2014})}\BibitemShut {NoStop}%
\bibitem [{\citenamefont {Sun}\ \emph {et~al.}(2009)\citenamefont {Sun},
  \citenamefont {Oeschler}, \citenamefont {Johnsen}, \citenamefont {Iversen},\
  and\ \citenamefont {Steglich}}]{sun_huge_2009}%
  \BibitemOpen
  \bibfield  {author} {\bibinfo {author} {\bibfnamefont {P.}~\bibnamefont
  {Sun}}, \bibinfo {author} {\bibfnamefont {N.}~\bibnamefont {Oeschler}},
  \bibinfo {author} {\bibfnamefont {S.}~\bibnamefont {Johnsen}}, \bibinfo
  {author} {\bibfnamefont {B.~B.}\ \bibnamefont {Iversen}}, \ and\ \bibinfo
  {author} {\bibfnamefont {F.}~\bibnamefont {Steglich}},\ }\href {\doibase
  10.1143/APEX.2.091102} {\bibfield  {journal} {\bibinfo  {journal} {Applied
  Physics Express}\ }\textbf {\bibinfo {volume} {2}},\ \bibinfo {pages}
  {091102} (\bibinfo {year} {2009})}\BibitemShut {NoStop}%
\bibitem [{\citenamefont {Buerger}(1932)}]{Buerger1932}%
  \BibitemOpen
  \bibfield  {author} {\bibinfo {author} {\bibfnamefont {M.~J.}\ \bibnamefont
  {Buerger}},\ }\href@noop {} {\bibfield  {journal} {\bibinfo  {journal} {Z.
  Kristallogr. -- Cryst. Mater.}\ }\textbf {\bibinfo {volume} {82}},\ \bibinfo
  {pages} {165} (\bibinfo {year} {1932})}\BibitemShut {NoStop}%
\bibitem [{\citenamefont {H{\"a}gg}(1928)}]{Hagg1928}%
  \BibitemOpen
  \bibfield  {author} {\bibinfo {author} {\bibfnamefont {G.}~\bibnamefont
  {H{\"a}gg}},\ }\href@noop {} {\bibfield  {journal} {\bibinfo  {journal} {Z.
  Kristallogr., Kristallgeom., Kristallphys., Kristallchem.}\ }\textbf
  {\bibinfo {volume} {68}},\ \bibinfo {pages} {470} (\bibinfo {year}
  {1928})}\BibitemShut {NoStop}%
\bibitem [{\citenamefont {Satterthwaite}\ and\ \citenamefont
  {Ure}(1957)}]{Satterthwaite1957}%
  \BibitemOpen
  \bibfield  {author} {\bibinfo {author} {\bibfnamefont {C.~B.}\ \bibnamefont
  {Satterthwaite}}\ and\ \bibinfo {author} {\bibfnamefont {R.~W.}\ \bibnamefont
  {Ure}},\ }\href {\doibase 10.1103/PhysRev.108.1164} {\bibfield  {journal}
  {\bibinfo  {journal} {Phys. Rev.}\ }\textbf {\bibinfo {volume} {108}},\
  \bibinfo {pages} {1164} (\bibinfo {year} {1957})}\BibitemShut {NoStop}%
\bibitem [{\citenamefont {Oganov}(2010)}]{oganov_modern_2010}%
  \BibitemOpen
  \bibfield  {author} {\bibinfo {author} {\bibfnamefont {A.~R.}\ \bibnamefont
  {Oganov}},\ }\href@noop {} {\emph {\bibinfo {title} {Modern {Methods} of
  {Crystal} {Structure} {Prediction}}}},\ \bibinfo {edition} {1st}\ ed.\
  (\bibinfo  {publisher} {Wiley-VCH Verlag GmbH \& Co. KGaA},\ \bibinfo {year}
  {2010})\BibitemShut {NoStop}%
\bibitem [{\citenamefont {Zhang}\ \emph {et~al.}(2017)\citenamefont {Zhang},
  \citenamefont {Wang}, \citenamefont {Lv},\ and\ \citenamefont
  {Ma}}]{zhang_materials_2017}%
  \BibitemOpen
  \bibfield  {author} {\bibinfo {author} {\bibfnamefont {L.}~\bibnamefont
  {Zhang}}, \bibinfo {author} {\bibfnamefont {Y.}~\bibnamefont {Wang}},
  \bibinfo {author} {\bibfnamefont {J.}~\bibnamefont {Lv}}, \ and\ \bibinfo
  {author} {\bibfnamefont {Y.}~\bibnamefont {Ma}},\ }\href {\doibase
  10.1038/natrevmats.2017.5} {\bibfield  {journal} {\bibinfo  {journal} {Nature
  Reviews Materials}\ }\textbf {\bibinfo {volume} {2}},\ \bibinfo {pages}
  {17005} (\bibinfo {year} {2017})}\BibitemShut {NoStop}%
\bibitem [{\citenamefont {Clarke}\ \emph {et~al.}(2016)\citenamefont {Clarke},
  \citenamefont {Walsh}, \citenamefont {Amsler}, \citenamefont {Malliakas},
  \citenamefont {Yu}, \citenamefont {Goedecker}, \citenamefont {Wang},
  \citenamefont {Wolverton},\ and\ \citenamefont {Freedman}}]{Clarke2016}%
  \BibitemOpen
  \bibfield  {author} {\bibinfo {author} {\bibfnamefont {S.~M.}\ \bibnamefont
  {Clarke}}, \bibinfo {author} {\bibfnamefont {J.~P.~S.}\ \bibnamefont
  {Walsh}}, \bibinfo {author} {\bibfnamefont {M.}~\bibnamefont {Amsler}},
  \bibinfo {author} {\bibfnamefont {C.~D.}\ \bibnamefont {Malliakas}}, \bibinfo
  {author} {\bibfnamefont {T.}~\bibnamefont {Yu}}, \bibinfo {author}
  {\bibfnamefont {S.}~\bibnamefont {Goedecker}}, \bibinfo {author}
  {\bibfnamefont {Y.}~\bibnamefont {Wang}}, \bibinfo {author} {\bibfnamefont
  {C.}~\bibnamefont {Wolverton}}, \ and\ \bibinfo {author} {\bibfnamefont
  {D.~E.}\ \bibnamefont {Freedman}},\ }\href {\doibase 10.1002/anie.201605902}
  {\bibfield  {journal} {\bibinfo  {journal} {Angew. Chem. Int. Ed.}\ }\textbf
  {\bibinfo {volume} {55}},\ \bibinfo {pages} {13446} (\bibinfo {year}
  {2016})}\BibitemShut {NoStop}%
\bibitem [{\citenamefont {Guo}\ \emph {et~al.}(2017)\citenamefont {Guo},
  \citenamefont {Akselrud}, \citenamefont {Bobnar}, \citenamefont {Burkhardt},
  \citenamefont {Schmidt}, \citenamefont {Zhao}, \citenamefont {Schwarz},\ and\
  \citenamefont {Grin}}]{guo_weak_2017}%
  \BibitemOpen
  \bibfield  {author} {\bibinfo {author} {\bibfnamefont {K.}~\bibnamefont
  {Guo}}, \bibinfo {author} {\bibfnamefont {L.}~\bibnamefont {Akselrud}},
  \bibinfo {author} {\bibfnamefont {M.}~\bibnamefont {Bobnar}}, \bibinfo
  {author} {\bibfnamefont {U.}~\bibnamefont {Burkhardt}}, \bibinfo {author}
  {\bibfnamefont {M.}~\bibnamefont {Schmidt}}, \bibinfo {author} {\bibfnamefont
  {J.-T.}\ \bibnamefont {Zhao}}, \bibinfo {author} {\bibfnamefont
  {U.}~\bibnamefont {Schwarz}}, \ and\ \bibinfo {author} {\bibfnamefont
  {Y.}~\bibnamefont {Grin}},\ }\href {\doibase 10.1002/anie.201700712}
  {\bibfield  {journal} {\bibinfo  {journal} {Angew. Chem. Int. Ed.}\ ,\
  \bibinfo {pages} {n/a}} (\bibinfo {year} {2017})}\BibitemShut {NoStop}%
\bibitem [{\citenamefont {Clarke}\ \emph {et~al.}(2017)\citenamefont {Clarke},
  \citenamefont {Amsler}, \citenamefont {Walsh}, \citenamefont {Yu},
  \citenamefont {Wang}, \citenamefont {Meng}, \citenamefont {Jacobsen},
  \citenamefont {Wolverton},\ and\ \citenamefont
  {Freedman}}]{clarke_creating_2017}%
  \BibitemOpen
  \bibfield  {author} {\bibinfo {author} {\bibfnamefont {S.~M.}\ \bibnamefont
  {Clarke}}, \bibinfo {author} {\bibfnamefont {M.}~\bibnamefont {Amsler}},
  \bibinfo {author} {\bibfnamefont {J.~P.~S.}\ \bibnamefont {Walsh}}, \bibinfo
  {author} {\bibfnamefont {T.}~\bibnamefont {Yu}}, \bibinfo {author}
  {\bibfnamefont {Y.}~\bibnamefont {Wang}}, \bibinfo {author} {\bibfnamefont
  {Y.}~\bibnamefont {Meng}}, \bibinfo {author} {\bibfnamefont {S.~D.}\
  \bibnamefont {Jacobsen}}, \bibinfo {author} {\bibfnamefont {C.}~\bibnamefont
  {Wolverton}}, \ and\ \bibinfo {author} {\bibfnamefont {D.~E.}\ \bibnamefont
  {Freedman}},\ }\href {\doibase 10.1021/acs.chemmater.7b01418} {\bibfield
  {journal} {\bibinfo  {journal} {Chem. Mater.}\ } (\bibinfo {year} {2017}),\
  10.1021/acs.chemmater.7b01418}\BibitemShut {NoStop}%
\bibitem [{\citenamefont {Amsler}\ \emph
  {et~al.}(2017{\natexlab{b}})\citenamefont {Amsler}, \citenamefont {Yao},\
  and\ \citenamefont {Wolverton}}]{amsler_cubine_2017}%
  \BibitemOpen
  \bibfield  {author} {\bibinfo {author} {\bibfnamefont {M.}~\bibnamefont
  {Amsler}}, \bibinfo {author} {\bibfnamefont {Z.}~\bibnamefont {Yao}}, \ and\
  \bibinfo {author} {\bibfnamefont {C.}~\bibnamefont {Wolverton}},\ }\href
  {http://arxiv.org/abs/1704.03038} {\bibfield  {journal} {\bibinfo  {journal}
  {arXiv:1704.03038 [cond-mat]}\ } (\bibinfo {year} {2017}{\natexlab{b}})},\
  \bibinfo {note} {arXiv: 1704.03038}\BibitemShut {NoStop}%
\bibitem [{\citenamefont {Naka}\ \emph {et~al.}(1976)\citenamefont {Naka},
  \citenamefont {Horii}, \citenamefont {Takeda},\ and\ \citenamefont
  {Hanawa}}]{naka_direct_1976}%
  \BibitemOpen
  \bibfield  {author} {\bibinfo {author} {\bibfnamefont {S.}~\bibnamefont
  {Naka}}, \bibinfo {author} {\bibfnamefont {K.}~\bibnamefont {Horii}},
  \bibinfo {author} {\bibfnamefont {Y.}~\bibnamefont {Takeda}}, \ and\ \bibinfo
  {author} {\bibfnamefont {T.}~\bibnamefont {Hanawa}},\ }\href {\doibase
  10.1038/259038a0} {\bibfield  {journal} {\bibinfo  {journal} {Nature}\
  }\textbf {\bibinfo {volume} {259}},\ \bibinfo {pages} {38} (\bibinfo {year}
  {1976})}\BibitemShut {NoStop}%
\bibitem [{\citenamefont {Goedecker}(2004)}]{goedecker2004}%
  \BibitemOpen
  \bibfield  {author} {\bibinfo {author} {\bibfnamefont {S.}~\bibnamefont
  {Goedecker}},\ }\href {\doibase 10.1063/1.3512900} {\bibfield  {journal}
  {\bibinfo  {journal} {J. Chem. Phys.}\ }\textbf {\bibinfo {volume} {120}},\
  \bibinfo {pages} {9911} (\bibinfo {year} {2004})}\BibitemShut {NoStop}%
\bibitem [{\citenamefont {Amsler}\ and\ \citenamefont
  {Goedecker}(2010)}]{amsler2010}%
  \BibitemOpen
  \bibfield  {author} {\bibinfo {author} {\bibfnamefont {M.}~\bibnamefont
  {Amsler}}\ and\ \bibinfo {author} {\bibfnamefont {S.}~\bibnamefont
  {Goedecker}},\ }\href {\doibase 10.1063/1.1724816} {\bibfield  {journal}
  {\bibinfo  {journal} {J. Chem. Phys.}\ }\textbf {\bibinfo {volume} {133}},\
  \bibinfo {pages} {224104} (\bibinfo {year} {2010})}\BibitemShut {NoStop}%
\bibitem [{\citenamefont {Amsler}\ \emph
  {et~al.}(2012{\natexlab{a}})\citenamefont {Amsler}, \citenamefont
  {{Flores-Livas}}, \citenamefont {Lehtovaara}, \citenamefont {Balima},
  \citenamefont {Ghasemi}, \citenamefont {Machon}, \citenamefont {Pailh\`{e}s},
  \citenamefont {Willand}, \citenamefont {Caliste}, \citenamefont {Botti},
  \citenamefont {San~Miguel}, \citenamefont {Goedecker},\ and\ \citenamefont
  {Marques}}]{amsler_crystal_2012}%
  \BibitemOpen
  \bibfield  {author} {\bibinfo {author} {\bibfnamefont {M.}~\bibnamefont
  {Amsler}}, \bibinfo {author} {\bibfnamefont {J.~A.}\ \bibnamefont
  {{Flores-Livas}}}, \bibinfo {author} {\bibfnamefont {L.}~\bibnamefont
  {Lehtovaara}}, \bibinfo {author} {\bibfnamefont {F.}~\bibnamefont {Balima}},
  \bibinfo {author} {\bibfnamefont {S.~A.}\ \bibnamefont {Ghasemi}}, \bibinfo
  {author} {\bibfnamefont {D.}~\bibnamefont {Machon}}, \bibinfo {author}
  {\bibfnamefont {S.}~\bibnamefont {Pailh\`{e}s}}, \bibinfo {author}
  {\bibfnamefont {A.}~\bibnamefont {Willand}}, \bibinfo {author} {\bibfnamefont
  {D.}~\bibnamefont {Caliste}}, \bibinfo {author} {\bibfnamefont
  {S.}~\bibnamefont {Botti}}, \bibinfo {author} {\bibfnamefont
  {A.}~\bibnamefont {San~Miguel}}, \bibinfo {author} {\bibfnamefont
  {S.}~\bibnamefont {Goedecker}}, \ and\ \bibinfo {author} {\bibfnamefont
  {M.~A.~L.}\ \bibnamefont {Marques}},\ }\href {\doibase
  10.1103/PhysRevLett.108.065501} {\bibfield  {journal} {\bibinfo  {journal}
  {Phys. Rev. Lett.}\ }\textbf {\bibinfo {volume} {108}},\ \bibinfo {pages}
  {065501} (\bibinfo {year} {2012}{\natexlab{a}})}\BibitemShut {NoStop}%
\bibitem [{\citenamefont {Flores-Livas}\ \emph {et~al.}(2012)\citenamefont
  {Flores-Livas}, \citenamefont {Amsler}, \citenamefont {Lenosky},
  \citenamefont {Lehtovaara}, \citenamefont {Botti}, \citenamefont {Marques},\
  and\ \citenamefont {Goedecker}}]{flores-livas_high-pressure_2012}%
  \BibitemOpen
  \bibfield  {author} {\bibinfo {author} {\bibfnamefont {J.~A.}\ \bibnamefont
  {Flores-Livas}}, \bibinfo {author} {\bibfnamefont {M.}~\bibnamefont
  {Amsler}}, \bibinfo {author} {\bibfnamefont {T.~J.}\ \bibnamefont {Lenosky}},
  \bibinfo {author} {\bibfnamefont {L.}~\bibnamefont {Lehtovaara}}, \bibinfo
  {author} {\bibfnamefont {S.}~\bibnamefont {Botti}}, \bibinfo {author}
  {\bibfnamefont {M.~A.~L.}\ \bibnamefont {Marques}}, \ and\ \bibinfo {author}
  {\bibfnamefont {S.}~\bibnamefont {Goedecker}},\ }\href {\doibase
  10.1103/PhysRevLett.108.117004} {\bibfield  {journal} {\bibinfo  {journal}
  {Phys. Rev. Lett.}\ }\textbf {\bibinfo {volume} {108}},\ \bibinfo {pages}
  {117004} (\bibinfo {year} {2012})}\BibitemShut {NoStop}%
\bibitem [{\citenamefont {Amsler}\ \emph
  {et~al.}(2012{\natexlab{b}})\citenamefont {Amsler}, \citenamefont
  {Flores-Livas}, \citenamefont {Huan}, \citenamefont {Botti}, \citenamefont
  {Marques},\ and\ \citenamefont {Goedecker}}]{amsler_novel_2012}%
  \BibitemOpen
  \bibfield  {author} {\bibinfo {author} {\bibfnamefont {M.}~\bibnamefont
  {Amsler}}, \bibinfo {author} {\bibfnamefont {J.~A.}\ \bibnamefont
  {Flores-Livas}}, \bibinfo {author} {\bibfnamefont {T.~D.}\ \bibnamefont
  {Huan}}, \bibinfo {author} {\bibfnamefont {S.}~\bibnamefont {Botti}},
  \bibinfo {author} {\bibfnamefont {M.~A.~L.}\ \bibnamefont {Marques}}, \ and\
  \bibinfo {author} {\bibfnamefont {S.}~\bibnamefont {Goedecker}},\ }\href
  {\doibase 10.1103/PhysRevLett.108.205505} {\bibfield  {journal} {\bibinfo
  {journal} {Phys. Rev. Lett.}\ }\textbf {\bibinfo {volume} {108}},\ \bibinfo
  {pages} {205505} (\bibinfo {year} {2012}{\natexlab{b}})}\BibitemShut
  {NoStop}%
\bibitem [{\citenamefont {Botti}\ \emph {et~al.}(2013)\citenamefont {Botti},
  \citenamefont {Amsler}, \citenamefont {Flores-Livas}, \citenamefont {Ceria},
  \citenamefont {Goedecker},\ and\ \citenamefont
  {Marques}}]{botti_carbon_2013}%
  \BibitemOpen
  \bibfield  {author} {\bibinfo {author} {\bibfnamefont {S.}~\bibnamefont
  {Botti}}, \bibinfo {author} {\bibfnamefont {M.}~\bibnamefont {Amsler}},
  \bibinfo {author} {\bibfnamefont {J.~A.}\ \bibnamefont {Flores-Livas}},
  \bibinfo {author} {\bibfnamefont {P.}~\bibnamefont {Ceria}}, \bibinfo
  {author} {\bibfnamefont {S.}~\bibnamefont {Goedecker}}, \ and\ \bibinfo
  {author} {\bibfnamefont {M.~A.~L.}\ \bibnamefont {Marques}},\ }\href
  {\doibase 10.1103/PhysRevB.88.014102} {\bibfield  {journal} {\bibinfo
  {journal} {Phys. Rev. B}\ }\textbf {\bibinfo {volume} {88}},\ \bibinfo
  {pages} {014102} (\bibinfo {year} {2013})}\BibitemShut {NoStop}%
\bibitem [{\citenamefont {Huan}\ \emph {et~al.}(2013)\citenamefont {Huan},
  \citenamefont {Amsler}, \citenamefont {Sabatini}, \citenamefont {Tuoc},
  \citenamefont {Le}, \citenamefont {Woods}, \citenamefont {Marzari},\ and\
  \citenamefont {Goedecker}}]{huan_thermodynamic_2013}%
  \BibitemOpen
  \bibfield  {author} {\bibinfo {author} {\bibfnamefont {T.~D.}\ \bibnamefont
  {Huan}}, \bibinfo {author} {\bibfnamefont {M.}~\bibnamefont {Amsler}},
  \bibinfo {author} {\bibfnamefont {R.}~\bibnamefont {Sabatini}}, \bibinfo
  {author} {\bibfnamefont {V.~N.}\ \bibnamefont {Tuoc}}, \bibinfo {author}
  {\bibfnamefont {N.~B.}\ \bibnamefont {Le}}, \bibinfo {author} {\bibfnamefont
  {L.~M.}\ \bibnamefont {Woods}}, \bibinfo {author} {\bibfnamefont
  {N.}~\bibnamefont {Marzari}}, \ and\ \bibinfo {author} {\bibfnamefont
  {S.}~\bibnamefont {Goedecker}},\ }\href {\doibase 10.1103/PhysRevB.88.024108}
  {\bibfield  {journal} {\bibinfo  {journal} {Phys. Rev. B}\ }\textbf {\bibinfo
  {volume} {88}},\ \bibinfo {pages} {024108} (\bibinfo {year}
  {2013})}\BibitemShut {NoStop}%
\bibitem [{\citenamefont {Roy}\ \emph {et~al.}(2008)\citenamefont {Roy},
  \citenamefont {Goedecker},\ and\ \citenamefont
  {Hellmann}}]{roy_bell-evans-polanyi_2008}%
  \BibitemOpen
  \bibfield  {author} {\bibinfo {author} {\bibfnamefont {S.}~\bibnamefont
  {Roy}}, \bibinfo {author} {\bibfnamefont {S.}~\bibnamefont {Goedecker}}, \
  and\ \bibinfo {author} {\bibfnamefont {V.}~\bibnamefont {Hellmann}},\ }\href
  {\doibase 10.1103/PhysRevE.77.056707} {\bibfield  {journal} {\bibinfo
  {journal} {Phys. Rev. E}\ }\textbf {\bibinfo {volume} {77}},\ \bibinfo
  {pages} {056707} (\bibinfo {year} {2008})}\BibitemShut {NoStop}%
\bibitem [{\citenamefont {Sicher}\ \emph {et~al.}(2011)\citenamefont {Sicher},
  \citenamefont {Mohr},\ and\ \citenamefont
  {Goedecker}}]{sicher_efficient_2011}%
  \BibitemOpen
  \bibfield  {author} {\bibinfo {author} {\bibfnamefont {M.}~\bibnamefont
  {Sicher}}, \bibinfo {author} {\bibfnamefont {S.}~\bibnamefont {Mohr}}, \ and\
  \bibinfo {author} {\bibfnamefont {S.}~\bibnamefont {Goedecker}},\ }\href
  {\doibase 10.1063/1.3530590} {\bibfield  {journal} {\bibinfo  {journal} {J.
  Chem. Phys.}\ }\textbf {\bibinfo {volume} {134}},\ \bibinfo {pages} {044106}
  (\bibinfo {year} {2011})}\BibitemShut {NoStop}%
\bibitem [{\citenamefont {Bl\"{o}chl}(1994)}]{PAW-Blochl-1994}%
  \BibitemOpen
  \bibfield  {author} {\bibinfo {author} {\bibfnamefont {P.~E.}\ \bibnamefont
  {Bl\"{o}chl}},\ }\href {\doibase 10.1103/PhysRevB.50.17953} {\bibfield
  {journal} {\bibinfo  {journal} {Phys. Rev. B}\ }\textbf {\bibinfo {volume}
  {50}},\ \bibinfo {pages} {17953} (\bibinfo {year} {1994})}\BibitemShut
  {NoStop}%
\bibitem [{\citenamefont {Kresse}(1995)}]{VASP-Kresse-1995}%
  \BibitemOpen
  \bibfield  {author} {\bibinfo {author} {\bibfnamefont {G.}~\bibnamefont
  {Kresse}},\ }\href {\doibase 10.1103/PhysRevB.47.558} {\bibfield  {journal}
  {\bibinfo  {journal} {J. Non-Cryst. Solids}\ }\textbf {\bibinfo {volume}
  {193}},\ \bibinfo {pages} {222} (\bibinfo {year} {1995})}\BibitemShut
  {NoStop}%
\bibitem [{\citenamefont {Kresse}\ and\ \citenamefont
  {Furthm\"{u}ller}(1996)}]{VASP-Kresse-1996}%
  \BibitemOpen
  \bibfield  {author} {\bibinfo {author} {\bibfnamefont {G.}~\bibnamefont
  {Kresse}}\ and\ \bibinfo {author} {\bibfnamefont {J.}~\bibnamefont
  {Furthm\"{u}ller}},\ }\href {\doibase 10.1016/0927-0256(96)00008-0}
  {\bibfield  {journal} {\bibinfo  {journal} {Comput. Mater. Sci.}\ }\textbf
  {\bibinfo {volume} {6}},\ \bibinfo {pages} {15} (\bibinfo {year}
  {1996})}\BibitemShut {NoStop}%
\bibitem [{\citenamefont {Kresse}\ and\ \citenamefont
  {Joubert}(1999)}]{VASP-Kresse-1999}%
  \BibitemOpen
  \bibfield  {author} {\bibinfo {author} {\bibfnamefont {G.}~\bibnamefont
  {Kresse}}\ and\ \bibinfo {author} {\bibfnamefont {D.}~\bibnamefont
  {Joubert}},\ }\href {\doibase 10.1103/PhysRevB.59.1758} {\bibfield  {journal}
  {\bibinfo  {journal} {Phys. Rev. B}\ }\textbf {\bibinfo {volume} {59}},\
  \bibinfo {pages} {1758} (\bibinfo {year} {1999})}\BibitemShut {NoStop}%
\bibitem [{\citenamefont {Perdew}\ \emph {et~al.}(1996)\citenamefont {Perdew},
  \citenamefont {Burke},\ and\ \citenamefont {Ernzerhof}}]{Perdew-PBE-1996}%
  \BibitemOpen
  \bibfield  {author} {\bibinfo {author} {\bibfnamefont {J.~P.}\ \bibnamefont
  {Perdew}}, \bibinfo {author} {\bibfnamefont {K.}~\bibnamefont {Burke}}, \
  and\ \bibinfo {author} {\bibfnamefont {M.}~\bibnamefont {Ernzerhof}},\ }\href
  {\doibase 10.1103/PhysRevLett.77.3865} {\bibfield  {journal} {\bibinfo
  {journal} {Phys. Rev. Lett.}\ }\textbf {\bibinfo {volume} {77}},\ \bibinfo
  {pages} {3865} (\bibinfo {year} {1996})}\BibitemShut {NoStop}%
\bibitem [{\citenamefont {Auer-Welsbach}\ \emph {et~al.}(1958)\citenamefont
  {Auer-Welsbach}, \citenamefont {Nowotny},\ and\ \citenamefont
  {Kohl}}]{auer-welsbach_untersuchung_1958}%
  \BibitemOpen
  \bibfield  {author} {\bibinfo {author} {\bibfnamefont {H.}~\bibnamefont
  {Auer-Welsbach}}, \bibinfo {author} {\bibfnamefont {H.}~\bibnamefont
  {Nowotny}}, \ and\ \bibinfo {author} {\bibfnamefont {A.}~\bibnamefont
  {Kohl}},\ }\href {\doibase 10.1007/BF00900636} {\bibfield  {journal}
  {\bibinfo  {journal} {Monatshefte für Chemie}\ }\textbf {\bibinfo {volume}
  {89}},\ \bibinfo {pages} {154} (\bibinfo {year} {1958})}\BibitemShut
  {NoStop}%
\bibitem [{\citenamefont {Jurriaanse}(2015)}]{jurriaanse_crystal_2015}%
  \BibitemOpen
  \bibfield  {author} {\bibinfo {author} {\bibfnamefont {T.}~\bibnamefont
  {Jurriaanse}},\ }\href {\doibase 10.1524/zkri.1935.90.1.322} {\bibfield
  {journal} {\bibinfo  {journal} {Zeitschrift für Kristallographie -
  Crystalline Materials}\ }\textbf {\bibinfo {volume} {90}},\ \bibinfo {pages}
  {322} (\bibinfo {year} {2015})}\BibitemShut {NoStop}%
\bibitem [{\citenamefont {Sun}\ \emph {et~al.}(2016)\citenamefont {Sun},
  \citenamefont {Dacek}, \citenamefont {Ong}, \citenamefont {Hautier},
  \citenamefont {Jain}, \citenamefont {Richards}, \citenamefont {Gamst},
  \citenamefont {Persson},\ and\ \citenamefont
  {Ceder}}]{sun_thermodynamic_2016}%
  \BibitemOpen
  \bibfield  {author} {\bibinfo {author} {\bibfnamefont {W.}~\bibnamefont
  {Sun}}, \bibinfo {author} {\bibfnamefont {S.~T.}\ \bibnamefont {Dacek}},
  \bibinfo {author} {\bibfnamefont {S.~P.}\ \bibnamefont {Ong}}, \bibinfo
  {author} {\bibfnamefont {G.}~\bibnamefont {Hautier}}, \bibinfo {author}
  {\bibfnamefont {A.}~\bibnamefont {Jain}}, \bibinfo {author} {\bibfnamefont
  {W.~D.}\ \bibnamefont {Richards}}, \bibinfo {author} {\bibfnamefont {A.~C.}\
  \bibnamefont {Gamst}}, \bibinfo {author} {\bibfnamefont {K.~A.}\ \bibnamefont
  {Persson}}, \ and\ \bibinfo {author} {\bibfnamefont {G.}~\bibnamefont
  {Ceder}},\ }\href {\doibase 10.1126/sciadv.1600225} {\bibfield  {journal}
  {\bibinfo  {journal} {Science Advances}\ }\textbf {\bibinfo {volume} {2}},\
  \bibinfo {pages} {e1600225} (\bibinfo {year} {2016})}\BibitemShut {NoStop}%
\bibitem [{\citenamefont {Wales}(2004)}]{wales_energy_2004}%
  \BibitemOpen
  \bibfield  {author} {\bibinfo {author} {\bibfnamefont {D.}~\bibnamefont
  {Wales}},\ }\href@noop {} {\emph {\bibinfo {title} {Energy {Landscapes}:
  {Applications} to {Clusters}, {Biomolecules} and {Glasses}}}},\ \bibinfo
  {edition} {1st}\ ed.\ (\bibinfo  {publisher} {Cambridge University Press},\
  \bibinfo {year} {2004})\BibitemShut {NoStop}%
\bibitem [{\citenamefont {Zhu}\ \emph {et~al.}(2014)\citenamefont {Zhu},
  \citenamefont {Liu}, \citenamefont {Pickard}, \citenamefont {Zou},\ and\
  \citenamefont {Ma}}]{zhu_reactions_2014}%
  \BibitemOpen
  \bibfield  {author} {\bibinfo {author} {\bibfnamefont {L.}~\bibnamefont
  {Zhu}}, \bibinfo {author} {\bibfnamefont {H.}~\bibnamefont {Liu}}, \bibinfo
  {author} {\bibfnamefont {C.~J.}\ \bibnamefont {Pickard}}, \bibinfo {author}
  {\bibfnamefont {G.}~\bibnamefont {Zou}}, \ and\ \bibinfo {author}
  {\bibfnamefont {Y.}~\bibnamefont {Ma}},\ }\href {\doibase 10.1038/nchem.1925}
  {\bibfield  {journal} {\bibinfo  {journal} {Nat Chem}\ }\textbf {\bibinfo
  {volume} {6}},\ \bibinfo {pages} {644} (\bibinfo {year} {2014})}\BibitemShut
  {NoStop}%
\bibitem [{\citenamefont {Giannozzi}\ \emph {et~al.}(2009)\citenamefont
  {Giannozzi}, \citenamefont {Baroni}, \citenamefont {Bonini}, \citenamefont
  {Calandra}, \citenamefont {Car}, \citenamefont {Cavazzoni}, \citenamefont
  {Ceresoli}, \citenamefont {Chiarotti}, \citenamefont {Cococcioni},
  \citenamefont {Dabo}, \citenamefont {Corso}, \citenamefont {de~Gironcoli},
  \citenamefont {Fabris}, \citenamefont {Fratesi}, \citenamefont {Gebauer},
  \citenamefont {Gerstmann}, \citenamefont {Gougoussis}, \citenamefont
  {Kokalj}, \citenamefont {Lazzeri}, \citenamefont {Martin-Samos},
  \citenamefont {Marzari}, \citenamefont {Mauri}, \citenamefont {Mazzarello},
  \citenamefont {Paolini}, \citenamefont {Pasquarello}, \citenamefont
  {Paulatto}, \citenamefont {Sbraccia}, \citenamefont {Scandolo}, \citenamefont
  {Sclauzero}, \citenamefont {Seitsonen}, \citenamefont {Smogunov},
  \citenamefont {Umari},\ and\ \citenamefont {Wentzcovitch}}]{espresso}%
  \BibitemOpen
  \bibfield  {author} {\bibinfo {author} {\bibfnamefont {P.}~\bibnamefont
  {Giannozzi}}, \bibinfo {author} {\bibfnamefont {S.}~\bibnamefont {Baroni}},
  \bibinfo {author} {\bibfnamefont {N.}~\bibnamefont {Bonini}}, \bibinfo
  {author} {\bibfnamefont {M.}~\bibnamefont {Calandra}}, \bibinfo {author}
  {\bibfnamefont {R.}~\bibnamefont {Car}}, \bibinfo {author} {\bibfnamefont
  {C.}~\bibnamefont {Cavazzoni}}, \bibinfo {author} {\bibfnamefont
  {D.}~\bibnamefont {Ceresoli}}, \bibinfo {author} {\bibfnamefont {G.~L.}\
  \bibnamefont {Chiarotti}}, \bibinfo {author} {\bibfnamefont {M.}~\bibnamefont
  {Cococcioni}}, \bibinfo {author} {\bibfnamefont {I.}~\bibnamefont {Dabo}},
  \bibinfo {author} {\bibfnamefont {A.~D.}\ \bibnamefont {Corso}}, \bibinfo
  {author} {\bibfnamefont {S.}~\bibnamefont {de~Gironcoli}}, \bibinfo {author}
  {\bibfnamefont {S.}~\bibnamefont {Fabris}}, \bibinfo {author} {\bibfnamefont
  {G.}~\bibnamefont {Fratesi}}, \bibinfo {author} {\bibfnamefont
  {R.}~\bibnamefont {Gebauer}}, \bibinfo {author} {\bibfnamefont
  {U.}~\bibnamefont {Gerstmann}}, \bibinfo {author} {\bibfnamefont
  {C.}~\bibnamefont {Gougoussis}}, \bibinfo {author} {\bibfnamefont
  {A.}~\bibnamefont {Kokalj}}, \bibinfo {author} {\bibfnamefont
  {M.}~\bibnamefont {Lazzeri}}, \bibinfo {author} {\bibfnamefont
  {L.}~\bibnamefont {Martin-Samos}}, \bibinfo {author} {\bibfnamefont
  {N.}~\bibnamefont {Marzari}}, \bibinfo {author} {\bibfnamefont
  {F.}~\bibnamefont {Mauri}}, \bibinfo {author} {\bibfnamefont
  {R.}~\bibnamefont {Mazzarello}}, \bibinfo {author} {\bibfnamefont
  {S.}~\bibnamefont {Paolini}}, \bibinfo {author} {\bibfnamefont
  {A.}~\bibnamefont {Pasquarello}}, \bibinfo {author} {\bibfnamefont
  {L.}~\bibnamefont {Paulatto}}, \bibinfo {author} {\bibfnamefont
  {C.}~\bibnamefont {Sbraccia}}, \bibinfo {author} {\bibfnamefont
  {S.}~\bibnamefont {Scandolo}}, \bibinfo {author} {\bibfnamefont
  {G.}~\bibnamefont {Sclauzero}}, \bibinfo {author} {\bibfnamefont {A.~P.}\
  \bibnamefont {Seitsonen}}, \bibinfo {author} {\bibfnamefont {A.}~\bibnamefont
  {Smogunov}}, \bibinfo {author} {\bibfnamefont {P.}~\bibnamefont {Umari}}, \
  and\ \bibinfo {author} {\bibfnamefont {R.~M.}\ \bibnamefont {Wentzcovitch}},\
  }\href {http://stacks.iop.org/0953-8984/21/i=39/a=395502} {\bibfield
  {journal} {\bibinfo  {journal} {J. Phys. Condens. Matter}\ }\textbf {\bibinfo
  {volume} {21}},\ \bibinfo {pages} {395502} (\bibinfo {year}
  {2009})}\BibitemShut {NoStop}%
\bibitem [{\citenamefont {He}\ \emph {et~al.}(2016)\citenamefont {He},
  \citenamefont {Amsler}, \citenamefont {Xia}, \citenamefont {Naghavi},
  \citenamefont {Hegde}, \citenamefont {Hao}, \citenamefont {Goedecker},
  \citenamefont {Ozolins},\ and\ \citenamefont
  {Wolverton}}]{he_ultralow_2016}%
  \BibitemOpen
  \bibfield  {author} {\bibinfo {author} {\bibfnamefont {J.}~\bibnamefont
  {He}}, \bibinfo {author} {\bibfnamefont {M.}~\bibnamefont {Amsler}}, \bibinfo
  {author} {\bibfnamefont {Y.}~\bibnamefont {Xia}}, \bibinfo {author}
  {\bibfnamefont {S.~S.}\ \bibnamefont {Naghavi}}, \bibinfo {author}
  {\bibfnamefont {V.~I.}\ \bibnamefont {Hegde}}, \bibinfo {author}
  {\bibfnamefont {S.}~\bibnamefont {Hao}}, \bibinfo {author} {\bibfnamefont
  {S.}~\bibnamefont {Goedecker}}, \bibinfo {author} {\bibfnamefont
  {V.}~\bibnamefont {Ozolins}}, \ and\ \bibinfo {author} {\bibfnamefont
  {C.}~\bibnamefont {Wolverton}},\ }\href {\doibase
  10.1103/PhysRevLett.117.046602} {\bibfield  {journal} {\bibinfo  {journal}
  {Phys. Rev. Lett.}\ }\textbf {\bibinfo {volume} {117}},\ \bibinfo {pages}
  {046602} (\bibinfo {year} {2016})}\BibitemShut {NoStop}%
\bibitem [{\citenamefont {Tse}\ \emph {et~al.}(2005)\citenamefont {Tse},
  \citenamefont {Iitaka}, \citenamefont {Kume}, \citenamefont {Shimizu},
  \citenamefont {Parlinski}, \citenamefont {Fukuoka},\ and\ \citenamefont
  {Yamanaka}}]{tse_electronic_2005}%
  \BibitemOpen
  \bibfield  {author} {\bibinfo {author} {\bibfnamefont {J.~S.}\ \bibnamefont
  {Tse}}, \bibinfo {author} {\bibfnamefont {T.}~\bibnamefont {Iitaka}},
  \bibinfo {author} {\bibfnamefont {T.}~\bibnamefont {Kume}}, \bibinfo {author}
  {\bibfnamefont {H.}~\bibnamefont {Shimizu}}, \bibinfo {author} {\bibfnamefont
  {K.}~\bibnamefont {Parlinski}}, \bibinfo {author} {\bibfnamefont
  {H.}~\bibnamefont {Fukuoka}}, \ and\ \bibinfo {author} {\bibfnamefont
  {S.}~\bibnamefont {Yamanaka}},\ }\href {\doibase 10.1103/PhysRevB.72.155441}
  {\bibfield  {journal} {\bibinfo  {journal} {Phys. Rev. B}\ }\textbf {\bibinfo
  {volume} {72}},\ \bibinfo {pages} {155441} (\bibinfo {year}
  {2005})}\BibitemShut {NoStop}%
\bibitem [{\citenamefont {Yamanaka}\ \emph {et~al.}(2000)\citenamefont
  {Yamanaka}, \citenamefont {Enishi}, \citenamefont {Fukuoka},\ and\
  \citenamefont {Yasukawa}}]{yamanaka_high-pressure_2000}%
  \BibitemOpen
  \bibfield  {author} {\bibinfo {author} {\bibfnamefont {S.}~\bibnamefont
  {Yamanaka}}, \bibinfo {author} {\bibfnamefont {E.}~\bibnamefont {Enishi}},
  \bibinfo {author} {\bibfnamefont {H.}~\bibnamefont {Fukuoka}}, \ and\
  \bibinfo {author} {\bibfnamefont {M.}~\bibnamefont {Yasukawa}},\ }\href
  {\doibase 10.1021/ic990778p} {\bibfield  {journal} {\bibinfo  {journal}
  {Inorg. Chem.}\ }\textbf {\bibinfo {volume} {39}},\ \bibinfo {pages} {56}
  (\bibinfo {year} {2000})}\BibitemShut {NoStop}%
\bibitem [{\citenamefont {Tanigaki}\ \emph {et~al.}(2003)\citenamefont
  {Tanigaki}, \citenamefont {Shimizu}, \citenamefont {Itoh}, \citenamefont
  {Teraoka}, \citenamefont {Moritomo},\ and\ \citenamefont
  {Yamanaka}}]{tanigaki_mechanism_2003}%
  \BibitemOpen
  \bibfield  {author} {\bibinfo {author} {\bibfnamefont {K.}~\bibnamefont
  {Tanigaki}}, \bibinfo {author} {\bibfnamefont {T.}~\bibnamefont {Shimizu}},
  \bibinfo {author} {\bibfnamefont {K.~M.}\ \bibnamefont {Itoh}}, \bibinfo
  {author} {\bibfnamefont {J.}~\bibnamefont {Teraoka}}, \bibinfo {author}
  {\bibfnamefont {Y.}~\bibnamefont {Moritomo}}, \ and\ \bibinfo {author}
  {\bibfnamefont {S.}~\bibnamefont {Yamanaka}},\ }\href {\doibase
  10.1038/nmat981} {\bibfield  {journal} {\bibinfo  {journal} {Nat Mater}\
  }\textbf {\bibinfo {volume} {2}},\ \bibinfo {pages} {653} (\bibinfo {year}
  {2003})}\BibitemShut {NoStop}%
\bibitem [{\citenamefont {Allen}\ and\ \citenamefont
  {Dynes}(1975)}]{Allen_1975}%
  \BibitemOpen
  \bibfield  {author} {\bibinfo {author} {\bibfnamefont {P.~B.}\ \bibnamefont
  {Allen}}\ and\ \bibinfo {author} {\bibfnamefont {R.~C.}\ \bibnamefont
  {Dynes}},\ }\href {\doibase 10.1103/PhysRevB.12.905} {\bibfield  {journal}
  {\bibinfo  {journal} {Phys. Rev. B}\ }\textbf {\bibinfo {volume} {12}},\
  \bibinfo {pages} {905} (\bibinfo {year} {1975})}\BibitemShut {NoStop}%
\end{thebibliography}
\providecommand{\noopsort}[1]{}\providecommand{\singleletter}[1]{#1}%

\end{document}